\documentstyle[12pt]{article}

\catcode`\@=11 \@addtoreset{equation}{section}\catcode`\@=12
\newcommand{\be}{\begin{equation}} \newcommand{\ee}{\end{equation}}
\newcommand{\ba}{\begin{eqnarray}} \newcommand{\ea}{\end{eqnarray}}

\begin{document}
\thispagestyle{empty}

\vskip 35mm

\begin{center}
{\large\bf The Mixmaster Cosmological Model as a Pseudo-Euclidean
Generalized Toda Chain}

\vskip 5mm
{\bf
Alexander Pavlov }\\
\vskip 5mm
     {\em Mathematical Simulation Institute\\
     Udmurt State University, 71 Krasnogeroyskaya St.,\\
     Izhevsk, 426034, Russia}\\
     e-mail:pavlov@matsim.udmurtia.su \\
\end{center}
\vskip 10mm

The question of the integrability of the mixmaster model of the Universe,
presented as a dynamical system with finite degrees of freedom, is
investigated in present paper. As far as the model belongs to the class of
pseudo-Euclidean generalized Toda chains \cite{Gavrilov}, the method of
getting Kovalevskaya exponents developed in \cite{KozlovTresh} for chains of
Euclidean type, is used. The generalized formula of Adler and van Moerbeke
\cite{Adler} for systems of an indefinite metric is obtained. There was shown
that although by the formula we got integer values of Kovalevskaya exponents
there were multivalued solutions, branched at particular points on a plane of
complex time $t$. This class of solutions differs from ones considered in
\cite{KozlovTresh}. Apparently, the system does not possess additional
algebraic \cite{Adler},\cite{Yoshida} and one-valued \cite{Ziglin} first
integrals because of complex and transcendental values of the exponents.
In addition, in the section 4 a ten-dimensional mixmaster model \cite{Stuckey}
was studied. There were not integer exponents.

\vskip 10mm


\vskip 30mm

\pagebreak


\section{The Pseudo-Euclidean Generalized Toda Chains}
\setcounter{equation}{0}

The mixmaster model belonges to pseudo-Euclidean generalized Toda chains
\cite{Gavrilov} on a level $H=0.$ The hamiltonian has a form
\begin{equation} 
H={1\over2}(-p_\alpha^2+p_{+}^2+p_{-}^2)+exp(4\alpha)V(\beta_{+},\beta_{-}),
\end{equation}
where the potential function $V(\beta_{+},\beta_{-})$ is an exponential
polinomial:
$$V(\beta_{+},\beta_{-})=exp(-8\beta_{+})+exp(4\beta_{+}+4\sqrt{3}\beta_{-})
+exp(4\beta_{+}-4\sqrt{3}\beta_{-})$$
\begin{equation} 
-2exp(4\beta_{+})-
2exp(-2\beta_{+}+2\sqrt{3}\beta_{-})-2exp(-2\beta_{+}-2\sqrt{3}\beta_{-}).
\end{equation}

The hamiltonian has a form:
\begin{equation} 
H={1\over2}<p,p>+\sum_{i=1}^N c_{i}v_{i},
\end{equation}
where $<,>$ is a scalar product in a Minkowski space $R^{1,n-1}$, $c_{i}$
are some real coefficients, $v_{i}\equiv exp(a_{i},q)$, $(,)$ is a scalar
product in a Euclid space $R^n$, $\vec a$ are real vectors. Pseudoeuclidity
of a momentum space is a distinctive peculiarity of gravitational problems
sothey can not be refered as analytical dynamics problems, where a quadratic
by momenta form is a kinetic energy.

On the other hand, the cosmological models can be considered as dynamical
systems. So it is possible to carry over strict methods of analysis,
traditionally used in the analytical mechanics, and adapt them to the
systems like (1.3). For analysis of integrability of the cosmological
model which is a hamiltonian system with three degrees of freedom, let us
apply the Painlev\'e test (see e.g. \cite{Bountis}) for calculation
of Kovalevskaya exponents \cite{Kov}. The term "Kovalevskaya exponents" was
introduced in paper \cite{Yoshida} thus it was marked an outstanding
contribution of the Russian woman into solution of the important problem
of dynamics-integration of a task of a rigid body rotation.
It was found that a general solution is meromorphic if an only if
the system possesses an additional one-valued \cite{Ziglin} or algebraic
\cite{Arh} first integral. Further development of the theory of
Kovalevskaya exponents is contained in \cite{Kozlov} \cite{Sadetov}.

By expanding $2n$-dimensional phase space to $2N$-dimensional one by
homeomorphism $(p,q)\mapsto (v,u)$:
\begin{equation} 
{v_{i}=exp(a_{i},q),\qquad u_{i}=<a_{i},p>,\qquad i=1,...,N,}
\end{equation}
generalizing the Flaska transformation \cite{Flaska}, one gets a hamiltonian
system whose movement equations are system of autonomic homogeneous
differential equations with polinomial right side:
$$\dot v_{i}=u_{i}v_{i},$$
\begin{equation} 
{\dot u_{i}=\sum_{j=1}^N M_{ij}v_{j},\qquad i=1,...,N.}
\end{equation}
The matrix $\hat M$ is constructed of scalar products of vectors $\vec a$ in
a Minkowski space $R^{1,n-1}$:
\begin{equation} 
M_{ij}\equiv -c_{j}<a_i,a_j>.
\end{equation}

The property of integrability of dynamical systems appears in a character of
singularities of solutions what it is not possible to say about points of
common position, so just singular points represent particular interest for
investigation. The equations (1.5) have the following partial solutions:
\begin{equation} 
{u_{i}=U_{i}/t,\qquad v_{i}=V_{i}/t^2,\qquad i=1,2,...,N,}
\end{equation}
coefficients $U_{i}$, $V_{i}$ obey a system of algebraic equations
$$-2V_i=U_iV_i,$$
\begin{equation} 
-U_i=\sum_{j=1}^N M_{ij}V_{j}.
\end{equation}
Now let us analize all types of solutions of the system (1.8) in detail.

I.Let $V_1\not= 0$, the rest $V_2,V_3,...,V_N=0$, then we get a solution:
if $M_{11}\not= 0$, then $V_1=2/M_{11}, U_1=-2, U_2=-2M_{21}/M_{11},...,
U_N=-2M_{N1}/M_{11}.$ Analogously the last solutions will be obtained. If for
some $i:$ $V_i\not= 0,$ and $V_j=0$ for all $j\not= i$, then if
$M_{ii}\not= 0$, and we get $U_i=-2, U_j=-2M_{ji}/M_{ii}$ for all $i\not= i$.
It follows from the obtained solutions that the significant point of analysis
is a nonequality of a correponding diagonal element of the matrix $\hat M$
to zero that is possible in case of an isotropy of a vector $a_i$. It is a
principial distinctive feature of pseudoeuclidean chains.

II.Let us pass to the next series of solutions. Let $V_1\not= 0$ and
$V_2\not= 0$, the other $V_i=0$, then $U_1=U_2=-2$, $V_1$ and $V_2$ are found
from a system
$$2=M_{11}V_1+M_{12}V_2,$$
\begin{equation} 
2=M_{21}V_1+M_{22}V_2.
\end{equation}
The condition of existence of solutions of the system (1.9) is a nondegeneracy
of a matrix $\hat M_{12}$. If $det\hat M_{12}\not= 0$, then, obtaining from
(1.9) values $V_1$ and $V_2$ and putting them in (1.8), one gets $U_3, U_4,...
,U_N:$
$$-U_3=M_{31}V_1+M_{32}V_2,$$
\begin{equation} 
-U_4=M_{41}V_1+M_{42}V_2,
\end{equation}
$$...$$
$$-U_N=M_{N1}V_1+M_{N2}V_2.$$
By the same way we obtain the rest of solutions of this series searching
through all possible nonzero pairs $V_i\not= 0$, $V_j\not=0$. The
corresponding condition of existence of solutions is a nondegeneracy of a
two-dimensional matrix: $det\hat M_{ij}\not= 0$, constructed of elements
$M_{ii}, M_{ij}, M_{ji}, M_{jj}.$

III.The following series of solutions are obtained by considering all nonzero
triple solutions $V_i, V_j, V_k$ with a condition of nondegeneracy of
corresponding three-dimensional matrices. The last possible solution is:
$U_i=-2, i=1,2,...,N, V_i=-(M_{ij})^{-1}U_j, det\hat M\not= 0$.

\section{A Calculation of Kovalevskaya Exponents}
\setcounter{equation}{0}

For investigation of a single-valuedness of obtained solutions we use Lyapunov
method \cite{KozlovTresh} based on studying of behaviour of solutions of
equations in variations. First variations obey the following differential
equations:
$${d\over dt}(\delta u)= \sum_{j=1}^N M_{ij}\delta v_j,$$
\begin{equation} 
{{d\over dt}(\delta v_i)={U_i\delta v_i\over t}+
{V_i\delta u_i\over t^2},\qquad i=1,...,N.}
\end{equation}
We seek their solutions in the form of
\begin{equation} 
{\delta u_i=\xi_i t^{\rho -1}, \qquad \delta v_i=\eta_i t^{\rho-2}, \qquad
i=1,...,N.}
\end{equation}
Then for searching of coefficients $\xi_i$, $\eta_i$ one gets a linear
homogeneous system of equations with a parameter $\rho$:
$$(\rho-2-U_i)\eta_i=V_i\xi_i,$$
\begin{equation} 
{(\rho-1)\xi_i=\sum_{j=1}^N M_{ij}\eta_j, \qquad i=1,...,N.}
\end{equation}
Values of a parameter $\rho$ are called Kovalevskaya exponents.

I.Let us consider solutions of the first series when $V_i\not= 0$. Let
$\eta_1\not= 0$ and the rest $\eta_i=0$, then from the first system of
equations (2.3) one gets $\xi_1 =M_{11}\rho\eta_1/2$, a substitution of
it to the second system (2.3) gives a condition on values of a parameter
$\rho$:$\rho (\rho -1)-2=0$, i.e. $\rho_1=-1$, $\rho_2=2$.
The remaining equation
(2.3) gives us solutions $\xi_i=\xi_i(\eta_1,\rho).$

Let then $\eta_2\not= 0$, then $\eta_3, \eta_4,...,\eta_N=0,
\rho=2-2M_{21}/M_{11}, \rho\eta_1=2\xi_1/M_{11}.$ The second system gets
functions $\xi_i=\xi_i(\eta_2), i=1,2,...,N.$

Further, considering $\eta_3\not= 0$, we receive $\rho=2-2<a_3,a_1>/<a_1,a_1>;
\eta_2, \eta_4,...,\eta_N=0; \rho\eta_1=2\xi_1/M_{11};$ and from the second
system: $\xi_i=\xi_i(\eta_2), i=1,2,...,N.$

As a result, having looked over all solutions of the first series for a case
$V_1\not= 0$, we obtain a formula for a spectrum $\rho:$
$\rho=2-2<a_i,a_1>/<a_1,a_1>, i=2,3,...,N.$

As a final result, having considered the rest solutions of the first series,
we obtain a formula for Kovalevskaya exponents $\rho$ that is the formula
of Adler and van Moerbeke for the case of indefinite spaces:
\begin{equation} 
{\rho=2-2{<a_i,a_k>\over <a_k,a_k>},\qquad i\not=k, \qquad
<a_k,a_k>\not=0.}
\end{equation}
The requirement $\rho \in Z$ is a necessary condition of a meromorphity of
solutions on a complex plane of $t$. It should be noticed that under
derivation of the formula (2.4) no restriction on a metric signature was
imposed. It is correct not only for spaces of the Minkowski signature.

II.The following series of solutions do not leed to the similar elegant
result. But we may show that solutions for variations exist in principle.
So, for example, let us take a solution of the second series when $V_1\ne 0,$
$V_2\ne 0$. Then we have:
$$\rho\eta_1=V_1\xi_1,$$
$$\rho\eta_2=V_2\xi_2,$$
\begin{equation} 
(\rho-2-U_3)\eta_3=0,
\end{equation}
$$...........$$
$$(\rho-2-U_N)\eta_N=0,$$
$${(\rho-1)\xi_i=\sum_{j=1}^N M_{ij}\eta_j,\qquad i=1,...,N.}$$
Let us assume, for instance, $\eta_3\ne 0,$ then $\eta_4,...,\eta_N=0,$
and therefore $\rho=2+U_3$. Out of the first pair of equations, expressing
$\eta_1=V_1\xi_1 /\rho,$ $\eta_2=V_2\xi_2 /\rho,$ we substitute them to the
second system. We obtain:
$$(\rho(\rho-1)-M_{11}V_1)\xi_1=M_{12}V_2\xi_2+\rho M_{13}\eta_3,$$
$$(\rho(\rho-1)-M_{21}V_1)\xi_1=M_{22}V_2\xi_2+\rho M_{23}\eta_3,$$
\begin{equation} 
\rho(\rho-1)\xi_3=M_{31}\xi_1+M_{32}\xi_2+M_{33}\rho\eta_3,
\end{equation}
$$.........$$
$$\rho(\rho-1)\xi_N=M_{N1}\xi_1+M_{N2}\xi_2+M_{N3}\rho\eta_3.$$
The first two equations represent a system of nonhomogeneous linear equations,
whence we get solutions $\xi_1=\xi_1(\eta_3),$ $\xi_2=\xi_2(\eta_3).$
Substituting them to the rest equations, we get $\xi_3=\xi_3(\eta_3),...,$
$\xi_N=\xi_N(\eta_3)$. By the same way the other solutions are to be obtained.

III. For the last solution when all $U_i=-2$ for $i=1,2,...,N,$ and $V_i$
are solutions of a matrix equation
\begin{equation} 
2=\sum_{j=1}^N M_{ij}V_j
\end{equation}
for the nondegenerate matrix $\hat M$, we receive an eigenvalue problem:
\begin{equation} 
\sum_{j=1}^N(M_{ij}V_j-\delta_{ij}\rho(\rho-1))\xi_j=0.
\end{equation}
As it is seen from (2.7), a sum of elements of any line of a matrix
$M_{ij}V_j$ is two. Out of this property of the matrix we obtain that one
eigenvector $\xi_i$ is unit with an eigenvalue $\rho(\rho-1)=2.$ From here
we obtain already known roots: $\rho=-1,2.$

\section{The Mixmaster Model of the Universe}
\setcounter{equation}{0}

Now let us apply the method to analysis of integrability of the mixmaster
model of the Universe, "root vectors" of which have a form (see (1.2)):
$${a_1(4,-8,0),\qquad a_4(4,4,0),}$$
\begin{equation} 
{a_2(4,4,4\sqrt{3}),\qquad a_5(4,-2,2\sqrt{3}),}
\end{equation}
$${a_3(4,4,-4\sqrt{3}),\qquad a_6(4,-2,-2\sqrt{3}).}$$
"Cartan matrix", composed of scalar products of the vectors (3.1) in
Minkowski space, then has a form:
\begin{equation} 
<a_i,a_j>=48\left(\matrix{\hfill 1&-1&-1&-1\hfill& 0& 0\cr
                          -1&\hfill 1&-1& 0& 0&-1\hfill\cr
                          -1&-1&\hfill 1& 0&-1\hfill& 0\cr
                          -1&\hfill 0&\hfill 0& 0&-1/2&-1/2\cr
                           \hfill 0&\hfill 0&-1&-1/2& 0&-1/2\cr
                           \hfill 0&-1&\hfill 0&-1/2&-1/2& 0\cr}\right).
\end{equation}

One gets, three "root vectors" dispose out of a light cone
(space-like vectors), the rest three are isotropic, are situated on the light
cone. Using the generalised Adler-van Moerbeke formula (2.4), taking account of
zero norm of three vectors, we get $\rho_1=2,$ $\rho_2=4,$ i.e. they are
integer.

Now we construct a matrix $\hat M$ by the formula (1.6), using a vector
$c_i(1,1,1,-2,-2,-2):$
\begin{equation} 
M_{ij}=48\left(\matrix{-1&\hfill 1&\hfill 1&-2&\hfill 0&\hfill 0\cr
                      \hfill 1&-1&\hfill 1&\hfill 0&\hfill 0&-2\cr
                      \hfill 1&\hfill 1&-1&\hfill 0&-2&\hfill 0\cr
                      \hfill 1&\hfill 0&\hfill 0&\hfill 0&-1&-1\cr
                      \hfill 0&\hfill 0&\hfill 1&-1&\hfill 0&-1\cr
                      \hfill 0&\hfill 1&\hfill 0&-1&-1&\hfill 0\cr}\right).
\end{equation}

By virtue of a specific form of the matrix (3.3), it is not difficult
a consideration of all partial solutions of the second type, when a pair
$V_i,V_j$ under some $i,j$ is not zero. One may notice that there is only one,
by virtue of nondegeneracy of blocks, nontrivial solution: $V_1\ne 0,$
$V_4\ne 0.$ It has a form: $V_1=1/24,$ $V_4=-1/24,$ $U_i=-2,$ $i=1,...,6.$

Equations on coefficients of variations of solutions are:
$$(\rho(\rho-1)+2)\eta_1+4\eta_4=0,$$
$$2\eta_1+\rho(\rho-1)\eta_4=0,$$
\begin{equation} 
(\rho-1)\xi_2=M_{21}\eta_1,
\end{equation}
$$(\rho-1)\xi_3=M_{31}\eta_1,$$
$$(\rho-1)\xi_5=M_{54}\eta_4,$$
$$(\rho-1)\xi_6=M_{64}\eta_4.$$

The condition of existence of solutions of the system of the first pair of
equations is a factorized algebraic equation of a fourth order on to a
spectral parameter $\rho:$
\begin{equation} 
[\rho(\rho-1)-2][\rho(\rho-1)+4]=0.
\end{equation}

The solutions of the first square equation are integer: $\rho_1=-1,$
$\rho_2=2,$ but roots of the second equation
\begin{equation} 
\rho_{1,2}=(1\pm i\sqrt{15})/2
\end{equation}
are complex and irrational. Thus, in a common case, a solution is not
one-valued on the complex plane of time $t$. Under some additional
conditions \cite{Adler},\cite{Yoshida},\cite{Ziglin}, it leads to
nonexistence additional algebraic or one-valued first integrals. Let us
remark that meromorphity of a common solution is a criterion of existence
of an algebraic first integral in dynamics of a rigid body (Gussont theorem)
and in a problem of three bodies (Brunce theorem).

\section{ A Mixmaster Model with Geometry $R^1\times S^3\times S^3\times S^3$}
\setcounter{equation}{0}

In paper \cite{Stuckey} a model of a spatially homogeneous and isotropic
vacuum universe with geometry $R^1\times S^3\times S^3\times S^3$ was studied.
A ten-dimensional model was chosen due to promising advances in a
ten-dimensional superstring theory. Its hamiltonian in Misner coordinates
$(\alpha,\beta_{\pm},\theta,\phi_{\pm},\eta,\psi_{\pm})$ is:
\begin{equation} 
H=-{p_\alpha^2\over 24}+\sum_{j=1}^8 {p_j^2\over 2}+{e^{16\alpha}\over 2}
[e^{-4\theta/{\sqrt{3}}}g(\beta)+e^{2\theta/{\sqrt{3}}}
(e^{-2\eta}g(\phi)+e^{2\eta}g(\psi))],
\end{equation}
where $g(x)$ means a function
\begin{equation} 
g(x)\equiv e^{4x_{+}+4\sqrt{3}x_{-}}+e^{4x_{+}-4\sqrt{3}x_{-}}+e^{-8x_{+}}-
2e^{4x_{+}}-2e^{2x_{+}+2\sqrt{3}x_{-}}-2e^{2x_{+}-2\sqrt{3}x_{-}}.
\end{equation}

The problem also belongs to the generalized pseudo-Euclidean Toda chains. Its
potential energy is a sum of 18 terms:
\begin{equation} 
U=\sum_{l=1}^{18} c_l exp(a_l,q).
\end{equation}
So 18 "root vectors" $\vec a$ in 9-dimensional Euclidean space are:
$$a_1(16,4,4\sqrt{3},-4/\sqrt{3},0,0,0,0,0),$$
$$a_2(16,4,-4\sqrt{3},-4/\sqrt{3},0,0,0,0,0),$$
$$a_3(16,-8,0,-4/\sqrt{3},0,0,0,0,0),$$
$$a_4(16,4,0,-4/\sqrt{3},0,0,0,0,0),$$
$$a_5(16,2,2\sqrt{3},-4/\sqrt{3},0,0,0,0,0),$$
$$a_6(16,2,-2\sqrt{3},-4/\sqrt{3},0,0,0,0,0),$$
$$a_7(16,0,0,2/\sqrt{3},4,4\sqrt{3},-2,0,0),$$
$$a_8(16,0,0,2/\sqrt{3},4,-4\sqrt{3},-2,0,0),$$
$$a_9(16,0,0,2/\sqrt{3},-8,0,-2,0,0_),$$
$$a_{10}(16,0,0,2/\sqrt{3},4,0,-2,0,0),$$
$$a_{11}(16,0,0,2/\sqrt{3},2,2\sqrt{3},-2,0,0),$$
$$a_{12}(16,0,0,2/\sqrt{3},2,-2\sqrt{3},-2,0,0),$$
$$a_{13}(16,0,0,2/\sqrt{3},0,0,2,4,4\sqrt{3}),$$
$$a_{14}(16,0,0,2/\sqrt{3},0,0,2,4,-4\sqrt{3}),$$
$$a_{15}(16,0,0,2/\sqrt{3},0,0,2,-8,0),$$
$$a_{16}(16,0,0,2/\sqrt{3},0,0,2,4,0),$$
$$a_{17}(16,0,0,2/\sqrt{3},0,0,2,2,2\sqrt{3}),$$
$$a_{18}(16,0,0,2/\sqrt{3},0,0,2,2,-2\sqrt{3}).$$

As before, we pass to redundant coordinates. Then get (1+8)-pseudo-Euclidean
space with metric $\eta_{ik}=diag(-1/12,1,...,1)$. By the same way as in
section 3 we calculate Kovalevskaya exponents using the Adler-van Moerbeke
formula (2.4). But besides integer numbers
$\rho=0,2,3,4$ there are rational exponents $\rho_1=4/3,$ $\rho_2=10/3.$
Apparently the considered model is nonintegrable.

\section{Discussion of Results}
\setcounter{equation}{0}

A search of first integrals in the form of algebraic and one-valued functions
is too restrictive. A more interesting question is about existence of
additional real-analytic integrals. Such a statement of the question goes back
to Poincar\'e. For Euclidean Toda chains a question about existence of
additional real-analytic polinomials by momenta first integrals was studied
in paper \cite{KT} where a classification of integrable systems was brought.

Furthermore,in this case, we are not interested in integrability at the all
phase space but at a concrete hypersurface "integral of energy" $H=0.$

The next step of investigation of the presence of chaotic characteristics is
a consideration of "truncated" model which to be obtain if one of the degrees
of freedom of the mixmaster model is fixed. I.e. it is obtained if we rewrite
a potential part of the hamiltonian (1.2), introducing coefficients in front
of its terms
$$V(\beta_{+},\beta_{-})=C_1 exp(-8\beta_{+})+
C_2 exp(4\beta_{+}+4\sqrt{3}\beta_{-})+
C_3 exp(4\beta_{+}-4\sqrt{3}\beta_{-})-$$
\begin{equation} 
2C_4 exp(4\beta_{+})-
2C_5 exp(-2\beta_{+}+2\sqrt{3}\beta_{-})-
2C_6 exp(-2\beta_{+}-2\sqrt{3}\beta_{-}),
\end{equation}
and put $C_2=C_3=C_5=C_6=0.$ Such a hamiltonian system in a four-dimensional
phase system can be studied with the help of the numerical analysis of
Poincar\'e mapping if its phase trajectories are situated in a compact region.
An existence of chaos in the "truncated" model implies
a stochasticity of the initial model. Let us remark that a calculation by
the generalized formula of Adler and van Moerbeke (2.4) for this case yields
integer $\rho$, as to additional exponents, they are also complex and
irrational, coincide with (3.6).

Because the Killing metric in the generalized Adler-van Moerbeke formula
is indefinite it should be pointed out at a classification scheme of
noncompact Lie algebras by analogy with using the Cartan scheme of
classification of compact simple Lie algebras for getting exact solutions
of Toda lattices as it was done in \cite{Bog}.

It is worth to mention paper \cite{Biesiada}, where was made an attempt of
proof of complicated behaviour of the mixmaster model trajectories out of an
analysis of level lines of the potential energy function.

Author is grateful to Dr. A.V.Borisov for fruitful numerous discussions and
acquintance with results of modern papers on dynamical systems.

\end{document}